\documentclass[epj]{svjour}
\usepackage{graphics}
\usepackage{graphicx}
\usepackage{amsmath}
  
\usepackage{amsthm}
\usepackage{amssymb}
\usepackage{cancel}
\usepackage{amsfonts}
\usepackage[utf8]{inputenc}
\usepackage{caption}
\usepackage{color}
\usepackage{enumerate}
\usepackage{subfigure}
\usepackage{textcomp}

\begin{document}

\title{Agent-based model for the h-index -- Exact solution}

\author{Barbara \.{Z}oga\l{}a-Siudem\inst{1,2,}%
\thanks{Corresponding author; e-mail: \texttt{zogala@ibspan.waw.pl}.}
 \and Grzegorz Siudem\inst{3}
 \and Anna Cena\inst{1,2}
 \and Marek Gagolewski\inst{1,4} 
}                     
%
%
\institute{Systems Research Institute, Polish Academy of Sciences,
ul. Newelska 6, 01-447 Warsaw, Poland
\and Institute of Computer Science, Polish Academy of Sciences, International PhD Studies Program
\and Faculty of Physics Warsaw University of Technology, ul. Koszykowa 75, 00-662 Warsaw, Poland
\and Faculty of Mathematics and Information Science, Warsaw University of Technology,\\
ul. Koszykowa 75, 00-662 Warsaw, Poland }
\date{Received: date / Revised version: date}

\abstract{
Hirsch's  $h$-index is perhaps the most popular citation-based measure
of scientific excellence. In 2013 G.~Ionescu and B.~Chopard proposed
an agent-based model describing a process for generating publications
and citations in an abstract scientific community.
Within such a framework, one may simulate a scientist's activity, and
-- by extension -- investigate the whole community of researchers.
Even though the Ionescu and Chopard model predicts the $h$-index quite well,
the authors provided a solution based solely on simulations. In this paper,
we complete their results with exact, analytic formulas. What is more,
by considering a simplified version of the Ionescu-Chopard model,
we obtained a compact, easy to compute formula for the $h$-index.
The derived approximate and exact solutions are investigated on
a simulated and real-world data sets.
%
} 
\maketitle
\section{Introduction}

Since the 1999 seminal paper by Barab\'{a}si and Albert \cite{BarabasiAlbert1999:preferential}
many methods that originally were developed in statistical physics
have been successfully applied in a wide range of problems coming from diverse domains.
Scientometrics, an area in which one is concerned with the quantitative characteristics of science
and scientific research, is one of such domains.
Recently, different authors studied -- among others -- the long term prediction
of scientific success \cite{Wang2013:Quantifying}, impact that an affiliation change
has on a scientist's productivity \cite{Deville2014:career}, or production and consumption
of the knowledge in physics \cite{ZhangetAll2013:physics,Mazloumian2013:foodweb}.
However, historically main efforts were focused on the study of
the structure of citation networks \cite{Redner1998:citationdist,Golosovsky2013:JStatPhys,Jeong,Eom2011:citationdynamisc},
and the reproduction of their degree distributions \cite{Eom2011:citationdynamisc,Krapivsky2000:network,Newman2011:clustering,GolosovskyPRL}.
Starting from the de Solla Price seminal work \cite{deSollaPrice1965} it is a known fact
that citation networks arise due to the preferential attachment rule \cite{BarabasiAlbert1999:preferential}.
This process, well known in complex network analysis
\cite{Jeong,Krapivsky2000:network,Newman2011:clustering}, was studied from
the point of view of citation networks \cite{Golosovsky2013:JStatPhys,Eom2011:citationdynamisc,GolosovskyPRL,WangPhysA1,WangPhysA2},
where it is also known as \textit{the rich get richer} rule or the Matthew effect \cite{Perc}.
Different variations of the classical, linear, preferential attachment
(see \cite{Krapivsky2000:network} or Table 1 in \cite{Golosovsky2013:JStatPhys})
were considered, but to the best of our knowledge there is a lack of models
in the literature which concern the $h$-index (except \cite{IonescuChopard}, which is
described in Sec. \ref{Sec:IC}).

The $h$-index proposed in 2005 by J.E.~Hirsch \cite{Hirsch2005:hindex}
is the most popular citation-based measure of scientific excellence.
Even though this data fusion tool was already studied in the 1940s
(compare the notion of the Ky Fan metric~\cite{Fan1943:metric}
and also the Sugeno integral, see, e.g., \cite{GagolewskiMesiar2014:Integrals}),
it may be conceived as a turning point in the history of scientometrics.
The idea standing behind the Hirsch index is to measure
not only the overall quality of a scientist's output
(most often expressed by the number of citations that each individual paper received),
but also its size. Thus, it may be understood as a measure of both
productivity and impact of a researcher (or an institution).
More formally, let us assume that we are given a list
$\mathbf{S}=(S_1,\dots, S_n)\in\mathbb{N}_0^n$, where $S_i$ denotes the number
of citations to the $i$-th paper.
If $S_{(n)}\ge 1$, the Hirsch index is given by the formula:
\[h\textrm{-index}=\max{\left\{ h=1,\dots, n:\
S_{(n-h+1)}\allowbreak\geq h\right\}},\]
where $S_{(n-h+1)}$ denotes the $(n-h+1)$-th order statistic of $\mathbf{S}$.
Moreover, if $S_{(n)}=0$, then $h \textrm{-index} = 0$.
Intuitively, an author has his/her $h$-index equal to $H$, if $H$ of her/his $n$ papers have
at least $H$ citations each, and the other $n-H$ papers have at most $H$ citations each.

There were a few papers devoted to the stochastic properties of the $h$-index
in some simple probabilistic models, see~\cite{EggheRousseau2006:informetrich,%
vanRaan2006:147chem,Burrell2007:hstochastic,Gagolewski2012:smps}).
Recently, Ionescu and Chopard in \cite{IonescuChopard} considered
a publication-citation process in an abstract scientific community
which was described by a multi-agent model.
Such a model consists of a scientist producing new papers, giving citations  to the already
published papers (including his/her own ones), and receiving citations from the community.
This bottom-up approach  allows to simulate a single scientist's activity as well as
to investigate the whole community of researchers.
What was very inspiring for us is the fact, that Fig. 3. in \cite{deSollaPrice1965}
is a perfect illustration of the mechanism of Ionescu-Chopard model,
but this de Solla Price article was
published almost $50$ years before Ionescu and Chopard paper.
Nevertheless, it turns out that their approach predicts quite well the
$h$-index from bibliometric data. However, its authors did not
provide an analytic form of a solution to their model, relying only on Monte Carlo
simulations instead. In the current work we present
an exact solution to that model as well as its simplification and an application on real-world data.

The paper is organized as follows. In Sec.~\ref{Sec:IC} the agent-based model proposed
by Ionescu and Chopard (referred to as the IC model) is described in very detail. Sec.~\ref{Sec:exact} presents
theoretical results concerning exact formulas for vectors of citations and the results of
comparative simulation studies. 
In Sec.~\ref{Sec:uproszczenie} a simplified model is proposed. Next, in
Sec.~\ref{Sec:emp} the results of an empirical analysis concerning all investigated approximations of
the $h$-index are presented. Finally, Sec.~\ref{Sec:conclusions} concludes the paper.

\section{The IC single-scientist model}\label{Sec:IC}

In 2013 Ionescu and Chopard \cite{IonescuChopard} introduced a multi-agent model
to describe a publication-citation generation process in an abstract scientific community.
Their approach consists of a scientist producing new papers, giving citations to his
own and other already published papers, and receiving citations from the community.
The model is based on a preferential attachment rule \cite{BarabasiAlbert1999:preferential},
which was observed in many real-world systems \cite{BarabasiAlbert1999:preferential,Jeong}.
As we mentioned before, preferential attachment rule is strongly connected
with the so-called Matthew effect \cite{Perc}: highly cited articles are
more eagerly cited by other authors than lowly cited ones.
More precisely,
the probability of adding new citations to a paper is proportional
to the number of citations it has already obtained.


\subsection{Simulation description}

Unlike in the case of various well-known models for constructing citation networks
\cite{Eom2011:citationdynamisc,GolosovskyPRL,WangPhysA1,WangPhysA2},
the IC model focuses not on the overall structure of a citation network
but only on the node degree distribution, 
i.e., on the number of citations of papers written by one author.
Its aim is to approximate citation scores for each published paper of
a given author, i.e.,~an $N$ dimensional vector $S=(S_1, \ldots, S_N)$,
where $S_k$ denotes the number of citations of the $k$-th paper.
By definition, this shall be based solely on the number $N$ of papers he/she
published as well as the total number $M$ of citations that his/her papers obtained.
Moreover, we assume that citations to each paper $S_k$ are of two kinds:
external $X_k$ and internal (self) ones $Y_k$, thus $S_k = X_k+Y_k$.

The simulation of interest is an iterative process.
We start with an initial number of papers $N_0$, none of which is cited.
During each iteration we add a new paper to the collection and distribute
both self and external citations to the existing papers according to the preferential
attachment rule. We give a fixed number of $p$ internal and $q$ external citations
to the $k$-th paper with probability of:
\begin{equation}\label{eq:pref}
p_k = \frac{X_k + 1}{\sum\limits_{l=1}^n X_l + n},\quad k=1,\ldots, n.
\end{equation}
Due to the form of the given probability distribution, in
\cite{IonescuChopard} it is assumed that only external citations
are taken  into account when assigning the new ones. Self citations do not
influence a paper's importance. Once the fixed number $N$ of published papers
is reached, the process goes on, but only $q$ external citations are
being granted during each step. The simulation ends as soon as
the total number of citations $M$ has been distributed.


\paragraph{Simulation steps in the IC model}
Let us now formalize the aforementioned procedure.
Such a detailed introduction is crucial for solving the model:
the simulation may end up on different stages depending on parameter values.
The IC model is based on the following input parameters:
\begin{enumerate}[(a)]
\item the number of papers $N\in\mathbb{N}$,
\item the total number of citations $M\in\mathbb{N}$,
\item the number of self citations added in each step $p\in\mathbb{N}$,
\item the number of external citations added in each step $q\in\mathbb{N}$ and
\item the initial number of papers with no citations at the beginning $N_0\in\mathbb{N}$.
\end{enumerate}
The initial values for sequences $\mathbf{X}$ and $\mathbf{Y}$ are given by
$X_1^{(0)} = 0, \ldots, X_N^{(0)} = 0$ and $Y_1^{(0)} = 0, \ldots, Y_N^{(0)} = 0$.
Values $X_{k}^{(t)}$ and $Y_{k}^{(t)}$ denote the number of external and self
citations, respectively, of the $k$-th paper in the $t$-th iteration.
Before the $k$-th paper is published, its citation counts are set to $0$.
Thus, $X_k^{(t)} = Y_k^{(t)} = 0$ for $k>t$. 
Nevertheless, please note that this assumption has no impact on further
derivations, as it is well-known that papers with no citations
do not influence the $h$-index value.

\bigskip
\noindent The simulation consists of the three following phases.
\paragraph{Phase 0.}
Firstly, we initialize the variables $X_1,\dots, X_{N_0}$ and $Y_1,\dots,\allowbreak Y_{N_0}$, and set $t=N_0$.
In the first step of the next phase we are going to distribute citations across
the first $N_0$ articles. In other words, the considered author has already
published her/his first $N_0$ articles and is waiting for citations.
Two cases are possible:
  \begin{itemize}
  \item $N_0 \geqslant N$ $\rightarrow$ the author published less than $N_0$ papers.
  In such a case, the simulation ends before going to phase (I),
  even though it is possible that there are still  citations left to be distributed.
  We could try going straight to phase (II) and distribute these citations,
  yet it would not increase the precision of the $h$-index estimation significantly
  (due to the fact that $N_0$ is small). On the other hand, this would unnecessarily
  complicate the formulas for $X_k$ and $Y_k$.
  \item $N_0 < N$ $\rightarrow$ there are enough papers and citations to go to phase (I).
  \end{itemize}
\paragraph{Phase (I)}
For each $t=N_0 + 1, \ldots, \min\{N, \lfloor \frac{M}{p+q} \rfloor + N_0 \}$,
we distribute  $q$ external and $p$ self citations according to the preferential
attachment rule given by Eq.~\eqref{eq:pref}:
\begin{align}
X_k^{(t)} &= X_k^{(t-1)} + \sum\limits_{j=0}^{q}j\cdot\mathbb{P}(X_k^{(t-1)}\rightarrow X_k^{(t-1)} + j)\label{eq:x_rec},\\
Y_k^{(t)} &= Y_k^{(t-1)} + \sum\limits_{j=0}^{p}j\cdot\mathbb{P}(Y_k^{(t-1)}\rightarrow Y_k^{(t-1)} + j)\label{eq:y_rec}.
\end{align}
When phase (I) comes to an end (which means that the author has already
published all her/his works and obtained all self citations), the three following cases are possible:
  \begin{itemize}
  \item $\frac{M}{p+q} + N_0 = N$ $\rightarrow\,$ simulation ends with no citations to distribute left,
  \item $\frac{M}{p+q} + N_0 < N$ $\rightarrow\,$ simulation ends, even if there are possibly
  up to $p+q-1$ undistributed citations left. In this case we could distribute such leftover
  citations, yet it would not increase the precision of the $h$-index estimation significantly and
  would unnecessarily complicate the formulas for $X_k$ and $Y_k$,
  \item $\frac{M}{p+q} + N_0 > N$ $\rightarrow\,$ simulation does not end, there are
  still citations   to be distributed. We go to phase (II).
  \end{itemize}
\paragraph{Phase (II)}
For each $t = N+1, \ldots, \lfloor \frac{M - (N-N_0)(p+q)}{q} \rfloor + N$,
we shall distribute only the external citations among the already published $N$ papers:
\begin{align}
X_k^{(t)} &= X_k^{(t-1)} + \sum\limits_{j=0}^{q}j\cdot\mathbb{P}(X_k^{(t-1)}\rightarrow X_k^{(t-1)} + j),\label{eq:xrec2}\\
Y_k^{(t)} &= Y_k^{(t-1)}.\nonumber
\end{align}
When phase (II) comes to an end, two situations are possible:
  \begin{itemize}
  \item $\left(M - (N-N_0)(p+q)\right)\ \mathrm{mod}\ q =0$ $\rightarrow\,$ simulation ends, no
  citations to distribute left,
  \item $\left(M - (N-N_0)(p+q)\right)\ \mathrm{mod}\ q \ne 0$ $\rightarrow\,$ simulation ends,
  even though there are possibly up to $q-1$ undistributed citations left.
  The reason to abandon the leftover citations distribution is the same as in phase (I).

  \end{itemize}


\section{Exact formulas for citation vectors}\label{Sec:exact}

Let us now present the exact formulas for $X_k^{(t)}$ and $Y_k^{(t)}$
derived for the IC model.


\subsection{External citations}\label{Sec:external}

Please notice that the sums in Eqs.~\eqref{eq:x_rec} and \eqref{eq:xrec2} are
in fact the expected values of random variables from
binomial distributions $\mathrm{Bin}(q, p_{k,t})$
and $\mathrm{Bin}(q, \tilde{p}_{k,t})$, respectively,
where probabilities $p_{k,t}$, $\tilde{p}_{k,t}$ are
given by:
\begin{align}
& p_{k, t} = \mathbb{P}(X_k^{(t-1)}\rightarrow X_k^{(t-1)} + 1) =\label{eq:pk}\\
=& \mathbb{P}(Y_k^{(t-1)}\rightarrow Y_k^{(t-1)} + 1)=
\begin{cases}
\frac{X_k^{(t-1)}+1}{\sum\limits_{l=1}^t X_l^{(t-1)} + t}, & k\leqslant t,\nonumber\\
0, & k>t,
\end{cases}
\end{align}
and
\begin{align}
\tilde{p}_{k, t} &= \mathbb{P}(X_k^{(t-1)}\rightarrow X_k^{(t-1)} + 1)\nonumber\\
& =
\frac{X_k^{(t-1)}+1}{\sum\limits_{l=1}^N X_l^{(t-1)} + N},\textrm{ for } k\leqslant N, \,t > N. \nonumber 
\end{align}

 The value of $X_k$ in the $t$-th step
can be written as:
\begin{align}\label{Eq:expectd-binomial}
X_k^{(t)} &=
\left\{ \begin{array}{ll}
X_k^{(t-1)} + qp_{k,t}, & t\leqslant N,\\
X_k^{(t-1)} + q\tilde{p}_{k,t}, & t > N,
\end{array}
\right. \\
&=\left\{ \begin{array}{ll}
X_k^{(t-1)} + \frac{q(X_k^{(t-1)}+1)}{\sum\limits_{l=1}^t X_l^{(t-1)} + t}, & t\leqslant N, \\
X_k^{(t-1)} + \frac{q(X_k^{(t-1)}+1)}{\sum\limits_{l=1}^N X_l^{(t-1)} + N}, & t > N. \nonumber
\end{array}
\right.
\end{align}
\noindent The sums in the denominators are equal to 
$$\sum\limits_{l=1}^{\min\{t,N\}} X_l^{(t-1)} = q(t-1-N_0).$$
Therefore,
\begin{align*}
X_k^{(t)} + 1 &=
\left\{ \begin{array}{ll}
(X_k^{(t-1)} + 1)\left(1+ \frac{q}{t(q+1) - q(N_0 + 1)}\right), & t\leqslant N, \\
(X_k^{(t-1)} + 1)\left(1 + \frac{q}{tq + N - q(N_0 + 1)}\right), & t > N,
\end{array}
\right.
\end{align*}
\noindent and now this recurrence relation can be solved easily. We wish to find $X_k = X_k^{(t_{\max})}$,
where the value of $t_{\max}$ depends on whether the simulation stops in phase (I) or (II). When solving
the recurrence equations we continue until reaching $X_k^{(k-1)}=0$ or $X_k^{(N_0)}=0$.
As a consequence, if $t_{\max}\leqslant N$, then we obtain:
\begin{equation*}
X_k = \prod\limits_{l=t_{\min}}^{t_{\max}}\left(1 + \frac{q}{l(q+1) - q(N_0 + 1)} \right) - 1,
\end{equation*}
and if $t_{\max} > N$, then it holds:
\begin{align*}
X_k =\allowbreak &\prod\limits_{l=t_{\min}}^{N}\left(1 + \frac{q}{l(q+1) - q(N_0 + 1)} \right)\times\\
\times&\prod\limits_{l=N+1}^{t_{\max}}\left(1 + \frac{q}{lq+N - q(N_0 + 1)} \right) - 1,
\end{align*}
\noindent where
\begin{align*}
t_{\min} &= \max\{N_0+1, k\},\\
t_{\max} &=
\left\{ \begin{array}{ll}
\lfloor \frac{M}{p+q} \rfloor + N_0 , & \textrm{ if } \lfloor \frac{M}{p+q} \rfloor + N_0 \leqslant N \label{eq:X_product},\\
\lfloor \frac{M - (N-N_0)(p+q)}{q} \rfloor + N_0 , & \textrm{ if } \lfloor \frac{M}{p+q} \rfloor + N_0 > N.
\end{array} \right.
\end{align*}
\noindent Please note that we can simplify the above formula
by relying on the notion of the $\Gamma$ function. Firstly, observe that:
\begin{align*}
\prod\left(1 + \frac{q}{l(q+1) - q(N_0 + 1)} \right) =
\frac{\prod\left(l - \alpha_1\right)}{\prod\left(l - \beta_1\right)},\\
\prod\left(1 + \frac{q}{lq + N - q(N_0 + 1)} \right) =
\frac{\prod\left(l - \alpha_2\right)}{\prod\left(l - \beta_2\right)},\\
\end{align*}
\noindent where:
\begin{align*}
&\alpha_1 = \frac{qN_0}{q+1},\quad \beta_1 = \frac{q(N_0+1)}{q+1}, \\
&\alpha_2 = \frac{qN_0-N}{q},\quad \beta_2 = \alpha_2+1.
\end{align*}
Moreover,
\begin{align*}
l-\alpha = \frac{\Gamma(l-\alpha+1)}{\Gamma(l-\alpha)},
\end{align*}
and
\begin{align*}
\prod\limits_{l=t_1}^{t_2}(l-\alpha) =
\frac{\Gamma(t_2-\alpha+1)}{\Gamma(t_1 - \alpha)}.
\end{align*}
Hence, the formula for $X_k$ can be written as:
\begin{equation}\label{eq:x_solved}
X_k = \frac{t_{\max}-\alpha_2}{N-\alpha_2}
\frac{\Gamma(N - \alpha_1 + 1)
\Gamma(t_{\min} - \beta_1)}
{\Gamma(t_{\min} - \alpha_1)
\Gamma(N - \beta_1+1)} - 1,
\end{equation}
with:
\begin{align*}
t_{\min} &= \max\{N_0 + 1, k\},\\
t_{\max} &=
\left\{ \begin{array}{ll}
\lfloor \frac{M}{p+q} \rfloor + N_0 , & \textrm{ if } \lfloor \frac{M}{p+q} \rfloor + N_0 \leqslant N, \\
\lfloor \frac{M - (N-N_0)(p+q)}{q} \rfloor + N_0 , & \textrm{ if } \lfloor \frac{M}{p+q} \rfloor + N_0 > N,
\end{array} \right.\\
\alpha_1 &= \frac{qN_0}{q+1},\quad \beta_1 = \frac{q(N_0+1)}{q+1}, \quad
\alpha_2 = \frac{qN_0-N}{q}.
\end{align*}
The above simplification gives a more elegant representation of $\mathbf{X}$.
However, it is worth noting that the product form
is more computationally stable than calculating gamma functions
for large arguments. Due to this fact in our simulations we use the product form.
Nevertheless, both representations enable us to compute the elements
of $\mathbf{X}$ significantly faster than in the case of the
simulation procedure presented in \cite{IonescuChopard}.

\subsection{Self citations}\label{Sec:self}

Similarly as in the previous subsection, we can solve the equation for self citations
distribution. Basing on Eqs.~\eqref{eq:y_rec} and \eqref{eq:pk},
we have:
\begin{small}
\begin{align*}
Y_k^{(t)} &= Y_k^{(t-1)} + pp_{k,t}\\
&=Y_k^{(t-1)} + \frac{p(X_k^{(t-1)}+1)}{q(t-1-N_0) + t}\\
&=Y_k^{(t-1)}\\
&\quad+ \frac{p}{q(t-1-N_0) + t}\prod\limits_{l=t_{\min}}^{t-1}\Big(1 + \frac{q}{q(l-1-N_0) + l} \Big)\\
&=\sum\limits_{i=t_{\min} + 1}^{t} \frac{p}{q(i-1-N_0) + i}\prod\limits_{l=t_{\min}}^{i-1}\Big(1 + \frac{q}{q(l-1-N_0) + l} \Big) \\
&\quad+ \frac{p}{(t_{\min}-1-N_0) + t_{\min}}.
\end{align*}
\end{small}%
\noindent%
We would like to find $Y_k = Y_k^{(s_{\max})}$, and due to the fact that here we
always end up in phase (I), $s_{\max}$ is equal to:
$$
s_{\max} = \min\{N, \lfloor \frac{M}{p+q} \rfloor + N_0\}.
$$
Hence,
\begin{small}
\begin{align*}
Y_k &=\sum\limits_{i=t_{\min}+1}^{s_{\max}} \frac{p}{q(i-1-N_0) + i}\prod\limits_{l=t_{\min}}^{i-1}\left(1 + \frac{q}{q(l-1-N_0) + l} \right) \\
&+ \frac{p}{(t_{\min}-1-N_0) + t_{\min}}.
\end{align*}
\end{small}%
Also, the formula for $Y_k$ may be simplified as follows:
\begin{align}
Y_k &=\sum\limits_{i=t_{\min}+1}^{s_{\max}} \frac{p}{q(i-1-N_0) + i} \frac{\Gamma(i-\alpha)\Gamma(t_{\min} -
\alpha)}{\Gamma(t_{\min} - \beta)\Gamma(i-\beta)}\nonumber\\
&+ \frac{p}{(t_{\min}-1-N_0) + t_{\min}},
\label{eq:y_solved}
\end{align}
where
$$
\alpha = \frac{qN_0}{q+1}, \quad \beta = \frac{q(N_0 + 1)}{q+1}.
$$

\subsection{Non-integer values of $p$ and $q$}

The authors of the IC model mention in \cite{IonescuChopard} that,
given non-integer values of $p$ and $q$, one distributes:
\begin{equation*}
 p'=\left\{\begin{array}{ll}
          \lceil p \rceil & \text{with probability }1-(\lceil p \rceil - p)\\
          \lceil p \rceil -1 & \text{with probability }(\lceil p \rceil -p)\\
          \end{array}\right.
\end{equation*}
self-citations as well as:
\begin{equation*}
 q'=\left\{\begin{array}{ll}
          \lceil q \rceil & \text{with probability } 1-(\lceil q \rceil -q)\\
          \lceil q \rceil -1 & \text{with probability }(\lceil q \rceil -q)\\
          \end{array}\right.
\end{equation*}
citations given by the scientific community.
Therefore, as with probabilities of $1-(\lceil p \rceil -p)$ and
$1-(\lceil q \rceil -q)$ the average number of self- and external citations is
equal to $p$ and $q$, respectively.

Let us consider $X_{k}^{(t)}$ and let
the second summand in Eq.~\eqref{Eq:expectd-binomial} be denoted as:
\begin{equation*}
\mathbb{E}(Q_{k}^{t})=\left\{\begin{array}{ll}
                              qp_{k,t} & t\leqslant N,\\
                              q\tilde{p}_{k,t} & t>N,\\
                             \end{array}\right.
\end{equation*}
where the number of external citations to the $k$-th paper obtained
at time $t$, $Q_{k}^{t}$ , follows the $\mathrm{Bin}(q, p_{k,t})$
distribution for $t\leqslant N$ and the $\mathrm{Bin}(q, \tilde{p}_{k,t})$ distribution for $t>N$.
Moreover,
\begin{equation*}
\mathbb{E}(Q_{k}^{t}|q')=\left\{\begin{array}{ll}
                                 \lceil q \rceil p_{k,t} & \text{with probability }1-(\lceil q \rceil- q),\\
				 (\lceil q \rceil -1)p_{k,t} & \text{with probability }(\lceil q \rceil- q).\\
                                \end{array}\right.
\end{equation*}
Taking into account the distribution of $q'$, we have that:
\begin{eqnarray*}
\mathbb{E}(Q_{k}^{t})&=&
\mathbb{E}(Q_{k}^{t}|q'=\lceil q \rceil)\,\mathbb{P}(q'=\lceil q \rceil)\\
&+&
\mathbb{E}(Q_{k}^{t}|q'=\lceil q \rceil-1)\,\mathbb{P}(q'=\lceil q \rceil-1).
\end{eqnarray*}
Therefore, Eq.~\eqref{Eq:expectd-binomial} is now of the form:
\begin{align*}
X_{k}^{(t)}&=X_k^{(t-1)} + \lceil q \rceil p_{k,t}(1+q-\lceil q \rceil)\\
&+(\lceil q \rceil -1) p_{k,t}(\lceil q \rceil- q)\\
&= X_k^{(t-1)} + (\lceil q \rceil + 1 + q - \lceil q \rceil -1)p_{k,t}\\
&=X_k^{(t-1)} + qp_{k,t}.
\end{align*}

By relying on a similar reasoning we obtain:
\begin{align*}
Y_{k}^{(t)}&=Y_k^{(t-1)} + \lceil p \rceil p_{k,t}(1+p-\lceil p \rceil)\\
&+(\lceil p \rceil -1) p_{k,t}(\lceil p \rceil- p)\\
&=Y_k^{(t-1)} + pp_{k, t}.
\end{align*}

It is easily seen that final form of the result is the same as for the formulas for integer $p$ and $q$.

\subsection{Overall number of citations and the h-index}

Once we have determined the formulas for external $X_k$ and self $Y_k$ citations
corresponding to the Ionescu and Chopard model, the only action left
to estimate $h$-index is just to sum them up.
One sees that both $X_k$ and $Y_k$ are nondecreasing, so $S_k = X_k + Y_k$ is also nondecreasing and thus:
$$
h_{\mathrm{exact}} = \max\{ k: S_k \geqslant k \}.
$$

\subsection{Comparative simulation study}\label{Sec:sim}

Let us now briefly compare the estimates of the $h\textrm{-index}$ obtained with the IC model (denoted as $h_{\mathrm{IC}}$)
and $h_{\mathrm{exact}}$, i.e., the ones that are based on Eq.~\eqref{eq:x_solved} and Eq.~\eqref{eq:y_solved}.
We consider the vector of citations of J.E.~Hirsch himself. The data were gathered
on July 30, 2015 from the Scopus database.
The vector consists of the total number of $M = 13480$ citations and
the total number of $N=205$ publications. The $h$-index of Hirsch is equal to $52$.

\begin{figure}[h!]
\centering
\includegraphics[width = 0.5\textwidth]{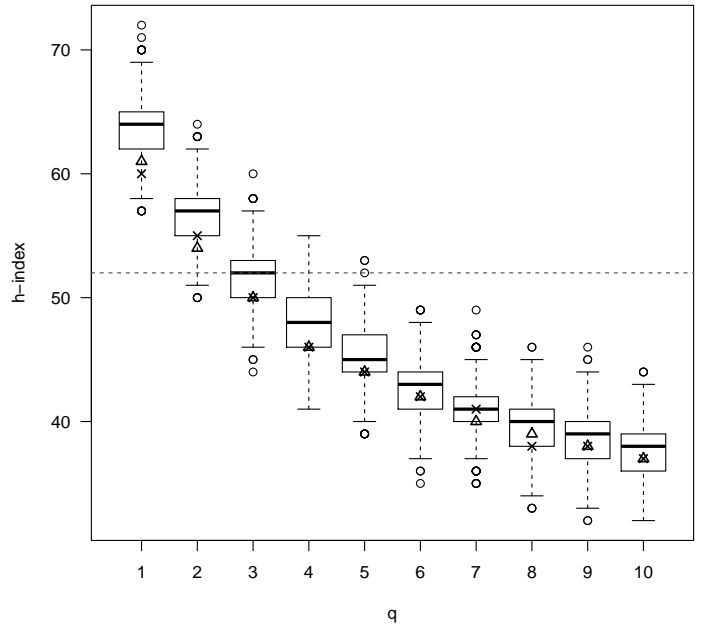}
\caption{Boxplots for the distribution of the $h$-index of Hirsch as estimated
via the IC model. Additionally, the $h$-index computed according
to citation vectors obtained via Eqs.~\eqref{eq:x_solved} and \eqref{eq:y_solved}
is marked with $\bigtriangleup$ and the $h$-index obtained from averaged citation vectors from the IC model
by $\boldsymbol\times$.}
\label{fig:boxplot}
\end{figure}

According to \cite{IonescuChopard}, parameters $p$ and $q$ giving
the best global agreement between the IC model and the original $h$-index
are equal to 1 and 2, respectively. However, the authors also stated that $p$ and $q$
can be tuned up in such a way that almost any scientific profile can be fit well.
In the case of the  $h$-index of Hirsch, we found out that $p=1$ and $q=3$
gives a reasonable agreement, while for $q=2$ the final $h$-index is overestimated.
Please note that the model is stochastic in its nature and its results vary across
different simulation runs, even for the same values of $p$, $q$, $M$ and $N$.
Therefore, for the purpose of a sensible comparison, we analyzed 1000 samples
for $p=1$, $q=1,2,\dots,10$ and $N_0=p+q$. The $h_{\mathrm{IC}}$ distribution estimates
are presented in Fig.~\ref{fig:boxplot} in a form of box-and-wiskers plots\footnote{%
The box-and-whisker plot aims to graphically represent an empirical distribution
of a given sample. The box ranges from the first ($Q_1$) to the third ($Q_3$) quartile
and the bold line gives the median. The whiskers range from $\max\{\mathrm{Min}, Q_1 - 1.5(Q_3-Q_1)\}$
to $\min\{\mathrm{Max}, Q_3 + 1.5(Q_3-Q_1)\}$. Moreover, each ($\circ$) marks an outlier,
that is an observation less than $Q_1 - 1.5(Q_3-Q_1)$ or greater than $Q_3 + 1.5(Q_3-Q_1)$).} Additionally,
the $h_{\mathrm{exact}}$
and the $h$-index obtained by averaging the citation vectors as generated by the IC model are indicated.
We may observe a high agreement between $h_{\mathrm{IC}}$ computed on
an averaged citation vector and $h_{\mathrm{exact}}$ (the largest difference
between these two estimates,~i.e., $|h_{\mathrm{IC}}-h_{\mathrm{exact}}|$ is equal to 1).

%

\begin{figure}
\centering
\subfigure[Vector of external citations $X$.]{\includegraphics[width = 0.4\textwidth]{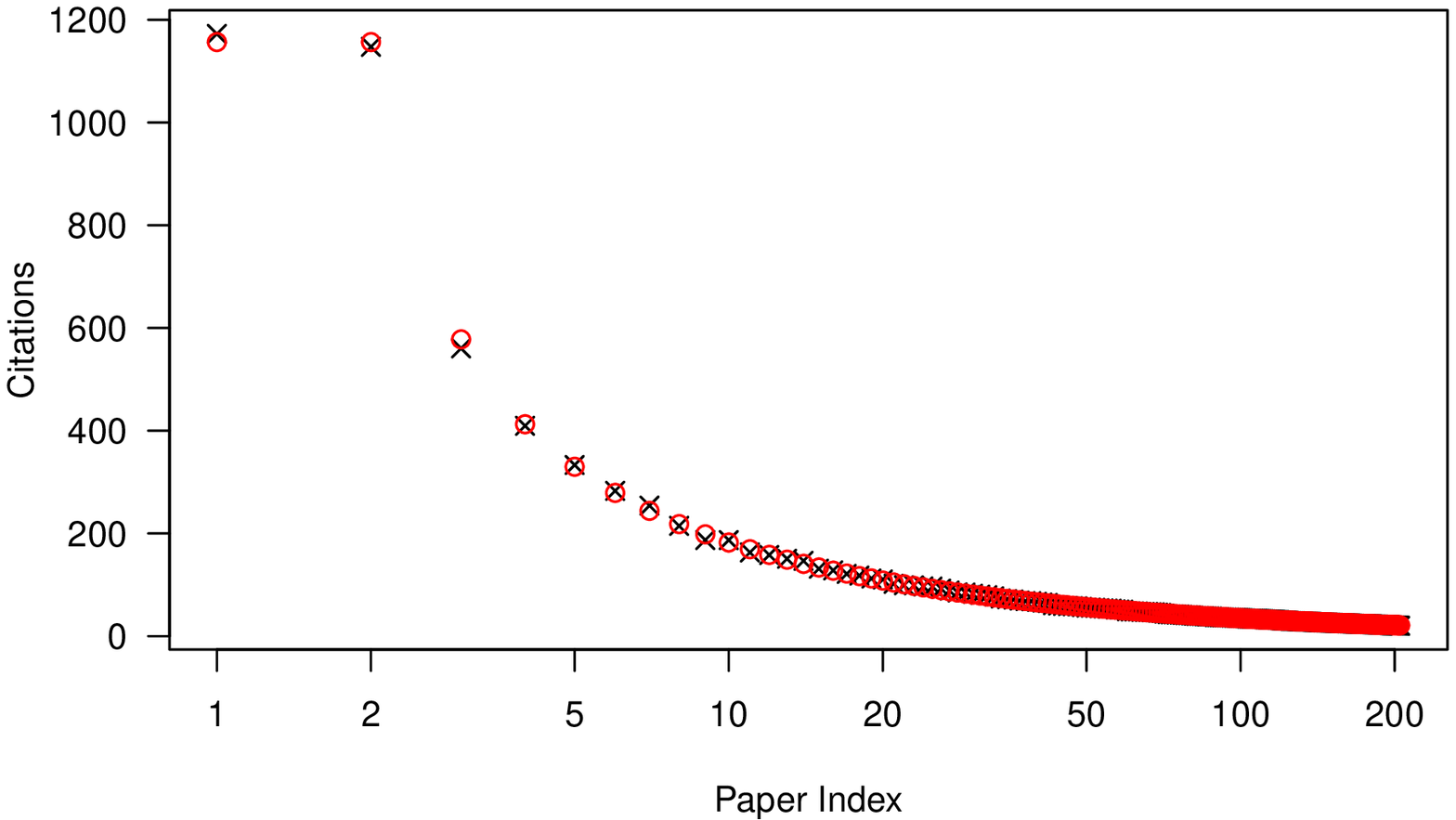}}
\subfigure[Vector of self citations $Y$.]{\includegraphics[width = 0.4\textwidth]{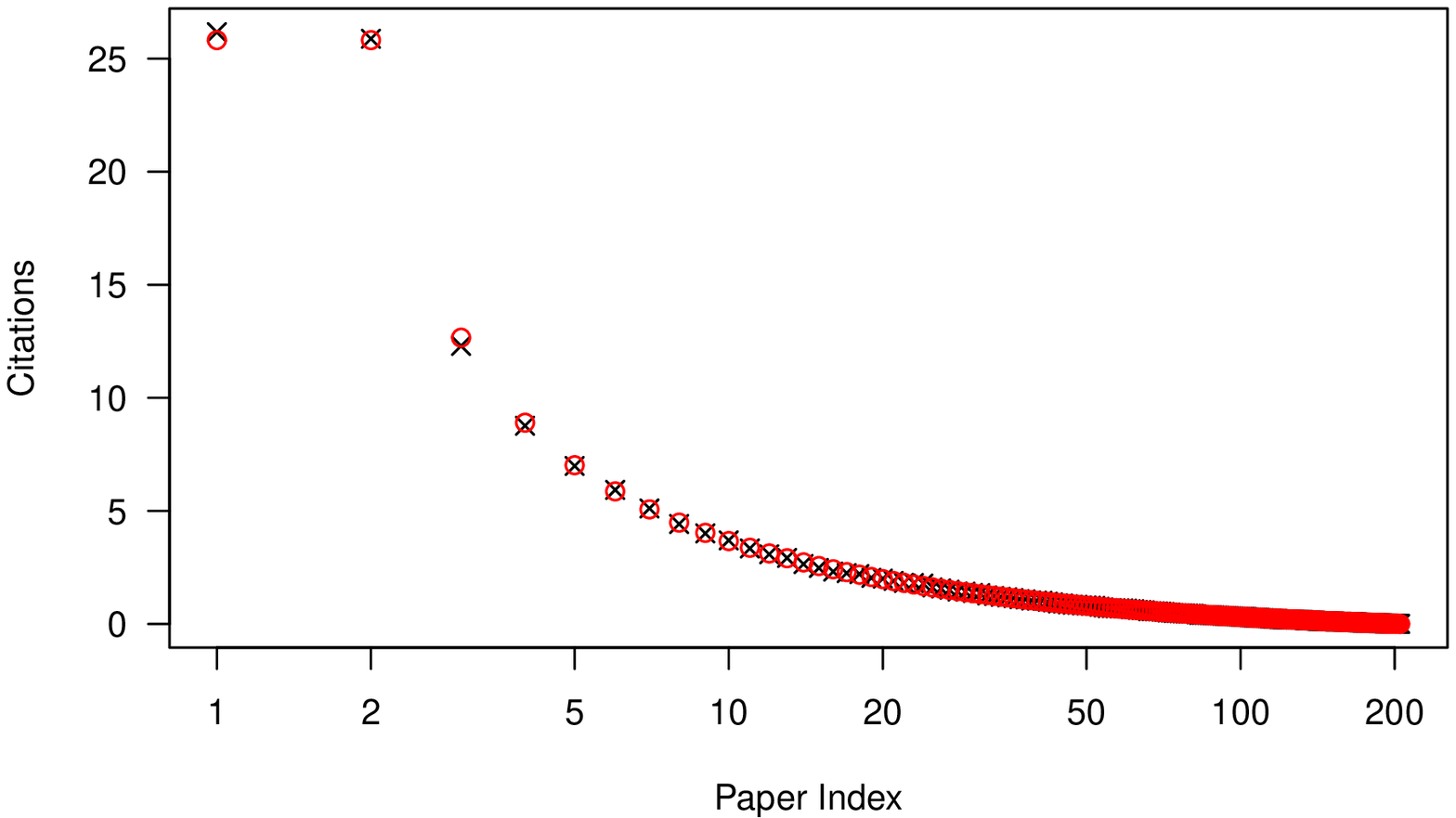}}
\caption{Step plots of vectors of external $X$ and self citations $Y$ obtained from the IC model,
depicted by $\boldsymbol\times$, and Eqs.~\eqref{eq:x_solved} and \eqref{eq:y_solved}, depicted
by \textcolor{red}{$\circ$}.}\label{Fig:XY}
\label{fig:XY}
\end{figure}

As it was stated in \cite{IonescuChopard}, the initial number of publications $N_0$ should be
small enough so as to not influence the rest of the process, but large enough in
order to provide enough papers to cite in the first iteration. The authors suggest to
choose $N_0 = p + q$. In order to assess the influence of this parameter
on $h_{\mathrm{IC}}$ and $h_{\mathrm{excat}}$, we
analyze $N_0$ varying from $1$ to $50$, for $p=1$ and $q=2$. Please note that
$N_0=50$ is nearly 25\% of all the Hirsch's publication count.
For the IC model, we perform 1000 runs and average the obtained values:
$\mathsf{AVR}(h_{\mathrm{IC}})$ denotes the mean of $h_{\mathrm{IC}}$
obtained in each run and $\mathrm{sd}$
its standard deviation, while $h_{\mathrm{IC}}(\mathsf{AVR})$ denotes the $h$-index
computed on an averaged citation vector from the IC model.
The results presented in Table~\ref{Tab:N0} suggest that there is no
significant difference between $N_0=1, 2$ and $N_0=p+q=3$ in this case.
Therefore, one may choose $N_0=1$ and if $N = 1$ and $M>0$,
simply assign the $h$-index equal to $1$.

\begin{table}[ht]
\centering
\caption{Aggregated results of 1000 runs of the IC model for $p=1$, $q=2$ and
$N_0\in\{1, \dots, 10, 15, 20, 25, 50\}$, where $\mathsf{AVR}(h_{\mathrm{IC}})$ denote
the mean $h_{\mathrm{IC}}$, sd its standard deviation, $h_{\mathrm{IC}}(\mathsf{AVR})$ the $h$-index
computed on an averaged citation vector from IC model and $h_{exact}$ -- the $h$-index
obtained via analytical formulas.}\label{Tab:N0}
\begin{tabular}{llllll}
  \hline
$N_0$ & \scriptsize{$\mathsf{AVR}(h_{\mathrm{IC}})$} & $sd$ & \scriptsize{$\mathsf{AVR}(h_{\mathrm{IC}})\pm sd$}
& \scriptsize{$h_{\mathrm{IC}}(\mathsf{AVR})$} & $h_{\mathrm{exact}}$ \\
  \hline
  \textbf{1} & \textbf{56.19} & 2.47 & (53.71;58.66) & \textbf{54} & \textbf{54} \\
  2 & 56.52 & 2.35 & (54.17;58.87) & 54 & 54 \\
  \textbf{3} & \textbf{56.73} & 2.36 & (54.38;59.09) & \textbf{54} & \textbf{54} \\
  4 & 57.04 & 2.38 & (54.66;59.42) & 55 & 55 \\
  5 & 57.34 & 2.30 & (55.04;59.64) & 55 & 55 \\
  6 & 57.51 & 2.30 & (55.21;59.81) & 55 & 55 \\
  7 & 57.81 & 2.37 & (55.44;60.18) & 56 & 56 \\
  8 & 58.21 & 2.27 & (55.94;60.48) & 57 & 56 \\
  9 & 58.43 & 2.34 & (56.08;60.77) & 56 & 56 \\
  10 & 58.63 & 2.43 & (56.2;61.07) & 57 & 56 \\
  15 & 59.70 & 2.33 & (57.36;62.03) & 58 & 58 \\
  20 & 60.85 & 2.29 & (58.56;63.14) & 60 & 60 \\
  25 & 61.79 & 2.30 & (59.49;64.09) & 61 & 62 \\
  50 & 65.23 & 2.30 & (62.93;67.53) & 71 & 71 \\
\end{tabular}
\end{table}

Fig.~\ref{Fig:XY} presents the step plots of vectors of external citations $\mathbf{X}$ (a) and
self citations $\mathbf{Y}$ (b) obtained from the averaged (over 1000 runs) IC model
as well as Eq.~\eqref{eq:x_solved}, \eqref{eq:y_solved}.
The sum of squared differences between the simulated and analytical results
are equal, respectively, 7.22 and 0.002.
The real value of the $h$-index  of Hirsch is equal to $52$ and the estimated values
(for parameters $p=1$ and $q=2$) are equal to
$h_{\mathrm{IC}}(\mathsf{AVR})=54$ and $h_{\mathrm{exact}}=54$.


\section{A simplification of the IC model}\label{Sec:uproszczenie}

Please note that the exact solution to the IC model, i.e., Eqs.~\eqref{eq:x_solved} and \eqref{eq:y_solved},
gives an analytical expression of the very intuitive and reasonable simulation setup
as proposed by Ionescu and Chopard. Nevertheless, as the form of the derived
formulas is quite complicated, their intuitive interpretation is difficult.

Let us recall that the IC model is based on an assumption that only external
citations are taken into account when assigning new ones (due to the form of
the probability distribution given by Eq.~\eqref{eq:pref}). Moreover,
during the simulation study, as it was also stated in~\cite{IonescuChopard},
we observed that the parameter $p$ has no significant influence on the outcoming $h$-index.
Hence, in this section we reduce the number of parameters, which leads to a
significant simplification of the model.

Let us employ the following assumptions:
\begin{enumerate}[(i)]
\item We assume $N_0=0$, so the first paper starts to gain citations just after its publication.
\item We consider only one vector $\mathbf{X}$, which means that we take into account
all the citations together without distinguishing between external and self citations.
\end{enumerate}
The number  of simulation parameters is decreased to only two: $q$, which is the
number of citations given in each iteration and $T$, which is number of simulation steps.
Similarly as in Sec.~\ref{Sec:exact} let us write the recurrence relation for $X_k^{(t)}$:
\begin{align*}
X_k^{(t)}&=X_k^{(t-1)} + \frac{q(X_k^{(t-1)}+1)}{\sum\limits_{l=1}^t X_l^{(t-1)} + t},\;\;\;&k=1,\,\dots,\,t,\\
X_k^{(t)}&=0,\;\;\;&k=t+1,\,\dots,\,T,
\end{align*}
which may be expressed as:
\begin{equation}
X_k=X_k(T)=\prod_{l=k}^{T}\frac{l}{l+q/(q+1)}-1,
\label{eq:X_approx1}
\end{equation}
which may be further simplified as:
\begin{equation}\label{eq:uproszczonyXk}
X_k=\frac{\Gamma(T+1)}{\Gamma(T+1-q/(q+1))}\frac{\Gamma(k-q/(q+1))}{\Gamma(k)}-1.
\end{equation}
Eq.~\eqref{eq:uproszczonyXk} is the exact solution of our simplified
version of the model, but the following asymptotic relation:
\begin{equation*}
\lim_{t\rightarrow\infty}\frac{\Gamma(t+\alpha)}{\Gamma(t)t^\alpha}=1,
\end{equation*}
allows us to obtain the approximation of $X_k$ as:
\begin{equation*} 
X_k\approx\frac{\Gamma(T+1)}{\Gamma(T+1)(T+1)^{-\alpha}}\frac{\Gamma(k)k^{-\alpha}}{\Gamma(k)}-1=\left(\frac{T+1}{k}\right)^\alpha-1,
\end{equation*}
where $\alpha=q/(q+1)$.

Even with our exact solution finding the compact formula
for the $h$-index seems untraceable. Fortunately, for the simplification
of the IC model, an observation that $X_k$ is an increasing function of $k$ leads to:
\begin{equation*}
h=\left(\frac{T+1}{h}\right)^\alpha-1,
\end{equation*}
which is equivalent to:
\begin{equation}
(h+1)h^\alpha=\left(T+1\right)^\alpha.
\label{eq:hirsch_eq}
\end{equation}
One can show that for every $T>0$ and $\alpha\in(0,1)$ Eq.~\eqref{eq:hirsch_eq} has always exactly one solution, which is the $h$-index.


\section{Real data evaluation}\label{Sec:emp}

In this section we perform an empirical analysis
of exemplary citation vectors gathered from Elsevier's Scopus
(see~\cite{Gagolewski2011:CITAN} for the detailed description of the data set).
Please note that the whole data set includes citation vectors corresponding to 16282 authors. Nevertheless,
about 78\% of all the vectors are of length one (among them ca.~32\%
represent a single uncited paper). This is typical to bibliometric data sets,
which consist of a high number of short vectors. Moreover, since it is
observed that all the vectors are skewed, usually to model them distributions
like exponential or Pareto type II (Lomax) are used,
(e.g., compare~\cite{BarczaTelcs2009:paretoimplyh,Glanzel2008:hconcatenation,Glanzel2008:statbibh}).
Table~\ref{Scopus:stats} presents basic sample statistics for the Scopus data set.

\begin{table}[ht]
\centering
\caption{Basic sample statistics (the Scopus data set) of the number of published papers by an author (N),
total number of citations he/she received (M), number of citations to his/her most (max)
and least (min) frequently cited paper.}\label{Scopus:stats}
\begin{tabular}{lllll}
  \hline
 & N & M & max & min \\
  \hline
  Min. & 1 & 0 & 0 & 0 \\
  1st Qu. & 1 & 0 & 0 & 0 \\
  Median & 1 & 3 & 3 & 1 \\
  Mean & 1.67 & 13.53 & 9.10 & 5.72 \\
  3rd Qu. & 1 & 11 & 9 & 5 \\
  Max. & 129 & 2396 & 836 & 836 \\
   \hline
\end{tabular}
\end{table}

For the sake of clarity of the results presented in this paper,
a subset of 100 randomly chosen authors has been selected. In order to assess
the quality of the proposed approximation we choose vectors of length greater than
or equal to 20 (in total number of 69) and from the vectors of length smaller than
20 we randomly choose 31 with uniform distribution. Basic sample statistics
of the selected sample are presented in
Table~\ref{Scopussample:stats}.

\begin{table}[ht]
\centering
\caption{Basic sample statistics
of the selected sample from the Scopus data set.}\label{Scopussample:stats}
\begin{tabular}{lllll}
  \hline
 & N & M & max & min \\
  \hline
  Min. & 1 & 0 & 0 & 0 \\
  1st Qu. & 6 & 45.75 & 19.50 & 0 \\
  Median & 21.50 & 207.50 & 36 & 0 \\
  Mean & 26.79 & 369.60 & 76.36 & 1.34 \\
  3rd Qu. & 31.75 & 486.20 & 102 & 0 \\
  Max. & 129 & 2396 & 636 & 20 \\
   \hline
\end{tabular}
\end{table}

\begin{figure}[h!]
\centering
\includegraphics[width = 0.5\textwidth]{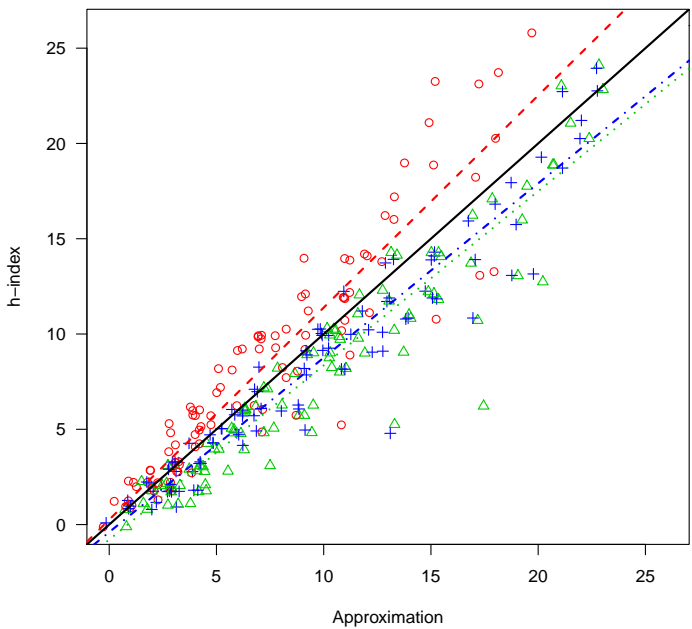}
\caption{Comparison of the $h$-index and its approximations on a Scopus data set.
The black continuous line is identity, so ideally
all the points should overlie this line.
The points depicted with \textcolor{red}{$\circ$} correspond to values
given by the $h$-index estimated from the exact solution of the IC simulation with parameters $N_0 = 1, p = 1, q' = 2$,
where only the vector of external citations was taken into account, 
i.e., one that is based on Eq.~\eqref{eq:x_solved}, 
the points marked with \textcolor{blue}{$+$} correspond to the estimate of the $h$-index that is
based only on a vector of external citations (Eq.~\eqref{eq:X_approx1}) with parameter $q''=3$,
while the points marked with \textcolor{green}{$\bigtriangleup$} correspond to the approximation given by  Eq.~\eqref{eq:hirsch_eq} with also $q''=3$.
The dotted lines of corresponding color, depicted as
\textcolor{red}{$\boldsymbol{---}$}, \textcolor{blue}{$\boldsymbol{-\cdot-\cdot-}$} and \textcolor{green}{$\boldsymbol{\cdot\cdot\cdot\cdot\cdot}$},
are the least
squares fit of the $h$-index values and considered approximations, respectively.
}
\label{fig:hirsch_approx}
\end{figure}

In Fig. \ref{fig:hirsch_approx} there are presented the approximated values of $h$-index as a function
of real values from considered data.
Please note that the mean squared difference between the estimated values and the $h$-index equals to
6.15, 6.45 and 4.14, respectively for the estimates based on Eq.~\eqref{eq:x_solved}, 
Eq.~\eqref{eq:hirsch_eq}
and Eq.~\eqref{eq:X_approx1}.

\begin{figure}[h!]
\centering
\includegraphics[width = 0.5\textwidth]{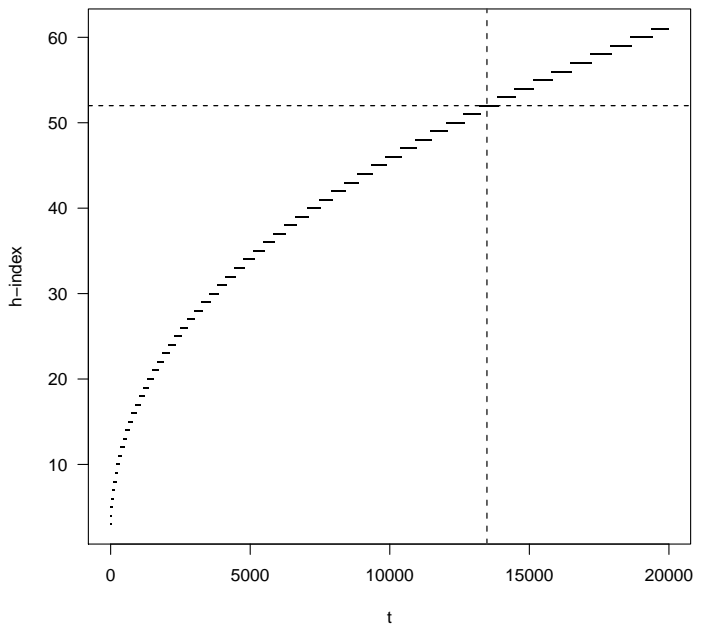}
\caption{The estimated values of the $h$-index of Hirsch himself based on the Eq.~\eqref{eq:X_approx1}
in each time point $t$ ($q=2.5$). The vertical line ($---$) depicts the real value of his $h$-index
(equal to 52), while the
vertical line depicts the current time point.}
\label{fig:prediction}
\end{figure}

Please note that in the case of Hirsch himself, considered in Sec.~\ref{Sec:sim},
the approximations of the $h$-index are equal to $51.99\approx52$ for approximation given by Eq.~\eqref{eq:hirsch_eq}
and $52$ for approximation based on Eq.~\eqref{eq:X_approx1}. The obtained estimates of the Hirsch $h$-index for
various values of parameter $q$ are presented in Table~\ref{Tab:Hirsch}. Please note that by an appropriate selection of the
parameter we were able to recreate the value of his $h$-index. Moreover, Fig.~\ref{fig:prediction}
depicts its predicted growth dynamic over each iteration.
We see that our simplification does not predict the $h$-index worse than the original simulation.
However, one should be aware that the approximate $h$-$\mathrm{index}$ given by Eq.~\eqref{eq:hirsch_eq}
is not necessarily an integer value (compare Fig.~\ref{fig:hirsch_approx} and Table~\ref{Tab:Hirsch}).
This should be taken into account in analysis of real data sets:
if needed, e.g., proper rounding can be applied.

\begin{table}[ht]
\centering
\caption{The approximations of
the Hirsch $h$-index calculated via Eq.~\eqref{eq:hirsch_eq} and based on Eq.~\eqref{eq:X_approx1}
for various parameters $q$.}\label{Tab:Hirsch}
\begin{tabular}{cccc}
  \hline
$q$ & Eq.~\eqref{eq:hirsch_eq} & Rounded values of Eq.~\eqref{eq:hirsch_eq} & Eq.~\eqref{eq:X_approx1} \\
  \hline
  1 & 23.14 & 23 & 23 \\
  1.5 & 34.75 & 35 & 35 \\
  2 & 44.27 & 44 & 44 \\
  \textbf{2.5} & \textbf{51.99} & \textbf{52} & \textbf{52} \\
  3 & 58.30 & 58 & 58 \\
  3.5 & 63.52 & 64 & 63 \\
  4 & 67.91 & 68 & 68 \\
  4.5 & 71.62 & 72 & 72 \\
  5 & 74.82 & 75 & 75 \\
   \hline
\end{tabular}
\end{table}

\section{Conclusions}\label{Sec:conclusions}
In this paper we investigated an agent-based model for
the bibliometric $h$-index introduced in~\cite{IonescuChopard}.
The main contribution included is an exact formula for the
number of external citations and self citations for each paper produced
by a given author. This result not only completes the work conducted by
Ionescu and Chopard, but also gives a perspective for a better insight
into the citation process. What is more, we proposed a simplification
of the IC model and presented the approximation of the $h$-index based
on such an approach. The obtained exact formulas were compared with
the results of simulations proposed by Ionescu and Chopard. Interestingly,
we may observe a good level  of compatibility between them, but mostly
for a large number of papers and citations. In this case, however, simulations
are more computationally demanding, which makes the usage of the exact formulas
more preferable. Also a real data evaluation on an informetric data set was presented.

There are still many issues worth deeper investigation.
First of all, due to the analytical formulas one may analyze the
theoretical properties of the $h$-$\mathrm{index}$ estimate. Since it
has been shown that the $h$-$\mathrm{index}$ is an example of an
aggregation operator and its properties can be studied by the means
of aggregation theory, it is worth to investigate
if such properties are still valid when it comes to the IC model estimate.

Moreover, also the theoretical evaluation of the influence of
the considered parameters on the results, which has been done by
Ionescu and Chopard only by an empirical study, should be performed.
Note that the exact formula for the approximation given by Eq.~(17) as well as the
comparative study of the proposed approximations of the Hirsch index and
the ones already available in the literature opens an interesting future research direction.
Also, it is reasonable to perform similar analysis on different data sets, for example
representing the data concerning the social network (Facebook, Twitter) users
or citation information gathered from different fields of science.

Also, there are a lot of variations of the classical preferential attachment rule, proposed by
Barab\'{a}si and Albert. There is also a rich discussion in the literature on the proper version of those
mechanisms  for considered problem \cite{Golosovsky2013:JStatPhys,Eom2011:citationdynamisc,Krapivsky2000:network}.
Mostly due to the simplicity (and for agreement with original work \cite{IonescuChopard}) we chose
a classical linear version \cite{Jeong,Newman2011:clustering}. The analysis
of different forms of preferential attachment rule (as those presented in \cite{GolosovskyPRL} or
\cite{Eom2011:citationdynamisc,Krapivsky2000:network}) is also left for future studies.


\section*{Acknowledgments}
The authors would like to thank the anonymous referees for all the useful comments
which helped to improve the quality of the manuscript.
Anna Cena and Barbara \.{Z}oga\l{}a-Siudem would like to acknowledge
the support by the European Union
from resources of the European Social Fund, Project PO KL
``Information technologies: Research and their interdisciplinary
applications'', agreement UDA-POKL.04.01.01-00-051/10-00
via the Interdisciplinary PhD Studies Program.
The study of Anna Cena and Marek Gagolewski
was partially supported by the National Science Center,
Poland, research project 2014/13/D/HS4/\-01700.

\section*{Author contributions}

Derived the results: BŻS, GS.
Analyzed the results: AC, MG.
Wrote the paper: BŻS, GS, AC, MG.







\end{document}